\documentclass[12pt,a4paper]{iopart}

\usepackage{graphicx}
\usepackage{color}
\usepackage{mathrsfs} 
\usepackage{dsfont}
\usepackage{iopams}  
\usepackage{tabularx}
\usepackage{multirow}
\usepackage[usenames,dvipsnames,table]{xcolor}
\usepackage{mathdots}
\usepackage[compress]{cite} 

\usepackage[bookmarks=true,
            bookmarksopen=true,colorlinks=false]{hyperref}

\newcommand{\eqref}{\eref}
\newcommand{\dd}{\mathrm{d}}

\newcommand{\R}{\mathds R}

\newcommand{\eps}{\varepsilon}

\newcommand{\Scri}{\mathscr{I}}

\newcommand{\Sp}{S_+}
\newcommand{\Sm}{S_-}
\newcommand{\Spm}{S_\pm}
\newcommand\diff[2]{\frac{\partial#1}{\partial#2}}
\newcommand{\dddot}[1]{\stackrel{\dots}{#1}}
\newcommand{\dilog}{\mathrm{dilog}}

\newcounter{mnotecount}

\newcommand{\mnotex}[1]
{\protect{\stepcounter{mnotecount}}$^{\mbox{\footnotesize $\bullet$\themnotecount}}$ 
\marginpar{
\raggedright\tiny\em
$\!\!\!\!\!\!\,\bullet$\themnotecount: #1} }

\begin{document}

\title[]{The characteristic initial value problem for the conformally invariant wave equation on a Schwarzschild background}

\author{J\"org Hennig}
\address{Department of Mathematics and Statistics,
           University of Otago,
           PO Box 56, \phantom{$^1$}Dunedin 9054, New Zealand}
\eads{\mailto{joerg.hennig@otago.ac.nz}}

\begin{abstract}
We resume former discussions of the conformally invariant wave equation on a Schwarzschild background, with a particular focus on the behaviour of solutions near the 'cylinder', i.e.\ Friedrich's representation of spacelike infinity. This analysis can be considered a toy model for the behaviour of the full Einstein equations 
and the resulting logarithmic singularities that appear to be characteristic for massive spacetimes. The investigation of the \emph{Cauchy} problem for the conformally invariant wave equation (Frauendiener and Hennig 2018, \emph{Class.\ Quantum\ Grav.} {\bf 35}  065015) showed that solutions generically develop logarithmic singularities at infinitely many expansion orders at the cylinder, but an arbitrary finite number of these singularities can be removed by appropriately restricting the initial data prescribed at $t=0$. From a physical point of view, any data at $t=0$ are determined from the earlier history of the system and hence not exactly `free data'. Therefore, it is appropriate to ask what happens if we `go further back in time' and prescribe initial data as early as possible, namely at a portion of past null infinity, and on a second past null hypersurface to complete the initial value problem. 
Will regular data at past null infinity automatically lead to a regular evolution up to future null infinity? Or does past regularity restrict the solutions too much, and regularity at both null infinities is mutually exclusive? Or do we still have suitable degrees of freedom for the data that can be chosen to influence regularity of the solutions to any desired degree?
In order to answer these questions, we study the corresponding \emph{characteristic} initial value problem.
In particular, we investigate in detail the appearance of singularities at expansion orders $n=0,\dots,4$ for angular modes $\ell=0,\dots,4$.
\\[2ex]{}
\end{abstract}


\section{Introduction}

The idealisation of isolated systems is a very valuable and fruitful concept
for the study of properties of individual configurations without the influence of the environment.  In general relativity, isolated systems correspond to asymptotically flat/simple spacetimes. These are most appropriately described in terms of Penrose's conformal compactification
\cite{Penrose1963, Penrose1964a, Penrose1964b},
which allows us to answer questions about global properties by investigating local properties at the conformal boundary. This boundary consists of the two null hypersurfaces past null infinity $\Scri^-$ and future null infinity $\Scri^+$, which approach each other at spacelike infinity $i^0$. Moreover,  $\Scri^-$ emanates from past timelike infinity $i^-$, and $\Scri^+$ focuses at future timelike infinity $i^+$.
For a comprehensive overview of conformal methods in general relativity, we refer to the well-known monographs \cite{Frauendiener2004,ValienteKroon2016}.

Clearly, the past and future timelike infinities $i^\pm$ are singular for any nonempty spacetime, since all matter and any other fields emanate/converge there. But even spacelike infinity $i^0$ is singular in general, which is a consequence of the gravitational field itself. Indeed, it is well-known that $i^0$ is regular only in very simple cases, like the Minkowski spacetime, but singular as soon as the ADM mass is nonzero --- even if the asymptotic region only contains vacuum. This was already observed by Penrose 
\cite{Penrose1965}. 

There is also a formulation of Einstein's field equations adapted to the setting of conformal compactifications, namely Friedrich's generalised conformal field equations
 \cite{Friedrich1998}. 
For a suitable treatment of spacelike infinity in this formalism, $i^0$ is represented as a cylinder $I$ of topology $S^2\times\R$. The cylinder and $\Scri^\pm$ approach each other at the critical sets $I^\pm$.

The conformal boundary is an important tool for the description of gravitational radiation. In particular, we can analyse the interaction between incoming and outgoing signals.
In a regular physical process, it is expected that incoming radiation from $\Scri^-$ interacts with fields in the spacetime (and with itself), and then eventually reaches $\Scri^+$, which is a physically reasonable interaction between $\Scri^-$ and $\Scri^+$. However, the equations do also allow signals from a point on $\Scri^-$ to exclusively travel within the conformal boundary: firstly along $\Scri^-$ to $I^-$, then 
further through the cylinder $I$, and finally via $I^+$ to $\Scri^+$.
Therefore, the cylinder is acting like a bridge across which information can travel from $\Scri^-$ to $\Scri^+$. Since the field equations become intrinsic transport equations on the cylinder, they do indeed completely determine the propagation of the field and all its derivatives at the cylinder. Such a process should be considered unphysical as it would not involve the physical spacetime at all. Hence, this was described as ``causality violation at infinity'' in 
\cite{BeyerFrauendienerHennig2020}.

The transport equations on the cylinder degenerate at the critical sets $I^\pm$.  As a consequence, the solutions tend to develop logarithmic singularities there 
\cite{ValienteKroon2002}
--- unless special initial data subject to certain regularity conditions are chosen. These singularities are expected to be further transported to $\Scri^+$, where they are superimposed to the physical information about outgoing radiation there. This effect is confirmed by the numerical studies in \cite{FrauendienerHennig2018,HennigMacedo2021}, and it provides another example of the above-mentioned causality violation at infinity.

In order to better understand these effects, the relation between the degree of regularity of solutions and certain fall-off conditions on the initial data has been studied for a number of different scenarios and equations.
Firstly, for the \emph{linearised Bianchi equations} on Minkowski \cite{Friedrich2003},  
a loss of smoothness of the solutions was confirmed, depending on how many regularity conditions are satisfied by the initial data. Corresponding numerical studies of the spin-2 equation on Minkowski can be found in
\cite{BeyerDoulis2012, BeyerDoulis2014, DoulisFrauendiener2013, MacedoValienteKroon2018}.

Similarly, the investigations of the \emph{Maxwell equations} on a Schwarzschild background in
\cite{ValienteKroon2007, ValienteKroon2009}
show that the electromagnetic field has logarithmic singularities, unless the initial data satisfy regularity conditions.

Furthermore, in a series of papers
\cite{FrauendienerHennig2014, FrauendienerHennig2017, FrauendienerHennig2018, HennigMacedo2021},
the behaviour of solutions to the \emph{conformally invariant wave equation} (i.e.\ the zero-rest-mass equation for spin zero) has been studied on Minkowski, Schwarzschild and Kerr backgrounds. 
This can be considered to be a toy model for more complicated problems, like the nonlinear stability of black holes.
Before we summarise the main observations from those publications, we also refer the reader to an interesting recent study that combines the efforts to understand the Maxwell equations and scalar fields: In 
\cite{MinucciMacedoKroon2022},
a nonlinear system of equations has been analysed, where the conformally invariant wave equation on Minkowski is coupled to an electromagnetic field via the covariant derivative operator. Again, logarithmic singularities are found, unless the initial data are appropriately fine-tuned. Moreover, the coupling of the two fields gives rise to additional singular terms, which disappear in the decoupled case.

We now come back to the conformally invariant wave equation on different background spacetimes as discussed in
\cite{FrauendienerHennig2014, FrauendienerHennig2017, FrauendienerHennig2018, HennigMacedo2021}. 
In the Minkowski case, it turned out that solutions are globally regular, provided a single regularity condition is satisfied by the initial data. For Schwarzschild and Kerr, on the other hand, only rather trivial solutions are regular everywhere. Generic solutions suffer from singularities at infinitely many orders. Consequently, one finds an entire hierarchy of conditions for regularity at different orders. By imposing appropriately many of these, any finite degree of regularity can be achieved.
We also note that an interesting additional feature of the Kerr background is a nonlinear coupling of angular modes in the solution. For example, for initial data containing only certain angular modes, other modes will also be excited in the time-evolution. This is not the case for a Schwarzschild background, where each mode evolves completely independently of the others.

Previous studies of the conformally invariant wave equation on Schwarzschild have focused on Cauchy problems with initial data at $t=0$. Here, we address the question about what happens if we go further back in time and provide initial data on a part of $\Scri^-$, which gives rise to a characteristic initial value problem. In particular, we are interested in the behaviour of solutions near $I^-$ and $I^+$, the type of singularities that can occur, and the relationship between the choice of characteristic initial data and the degree of regularity of the corresponding solutions.

The basis for the forthcoming analysis is to identify suitable well-behaved coordinates for the Schwarzschild solution that cover the cylinder as well as parts of future and past null infinity. This will be done in Sec.~\ref{sec:compactification}. 
We then study the behaviour of solutions to the conformally invariant wave equation at infinity in Sec.~\ref{sec:infinity}. To this end, we first analyse the solutions near the cylinder at spacelike infinity, and then also investigate the behaviour in a neighbourhood of $\Scri^-$. This will allow us to relate the asymptotic structure of the initial data to the behaviour on the cylinder. Afterwards, in Sec.~\ref{sec:Examples}, we apply the findings and study three families of initial data, corresponding to the three lowest angular modes. Finally, we discuss the results in Sec.~\ref{sec:discussion}.

\section{Conformal compactification}\label{sec:compactification}

\begin{figure}
 \centering
 \includegraphics[width=8cm]{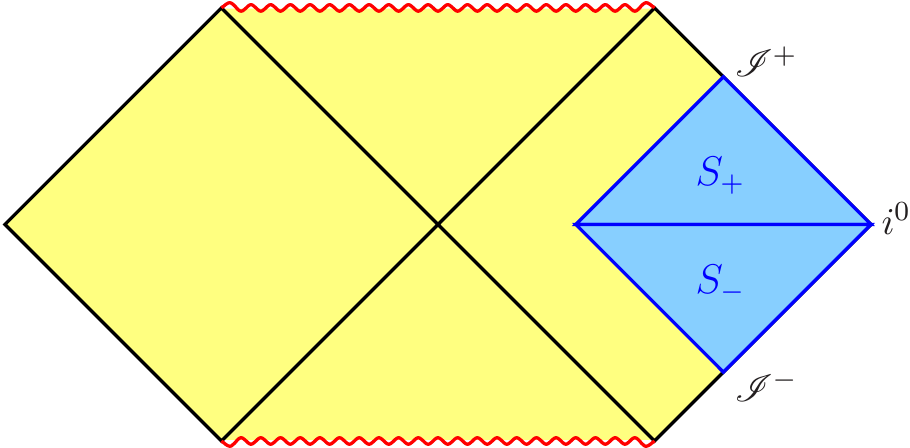}
 \caption{Conformal diagram of the Schwarzschild background. We introduce new coordinates in the blue region near spacelike infinity, with separate coordinate patches in $\Sp$ and $\Sm$.}
 \label{fig:ConfDiag}
\end{figure}

We consider a region near spacelike infinity in the Schwarzschild spacetime that is enclosed by lightlike boundaries, like the blue region shown in Fig.~\ref{fig:ConfDiag}. It would be possible to cover this entire region with just a single coordinate patch, for example using suitable modifications of Kruskal's extension. A particular coordinate system of this type is described in \cite{HalacekLedvinka}. While such coordinates may be well-behaved near null infinity $\Scri^\pm$, they usually introduce logarithmic terms into the metric as spacelike infinity $i^0$ is approached. Hence they are not optimal for a subsequent transformation that blows up $i^0$ to the cylinder $I$. However, this problem can easily be circumvented if we do not insist on using coordinates that extend to both $\Scri^+$ and $\Scri^-$, but instead choose separate coordinates in the regions $\Sp$ (which contains a part of $\Scri^+$) and $\Sm$ (which contains a part of $\Scri^-$), see Fig.~\ref{fig:ConfDiag}.

One possible coordinate choice in $\Sp$, which is due to Friedrich \cite{Friedrich2004}, was the basis for the analytical and numerical considerations in \cite{FrauendienerHennig2017, FrauendienerHennig2018}. These coordinates are constructed with a series of  transformations starting from the Schwarzschild metric in \emph{isotropic coordinates}. Another possibility is to apply similar coordinate transformations starting from \emph{Schwarzschild coordinates}. A generalisation of this idea was employed in \cite{HennigMacedo2021}, where a conformal compactification for a part of the Kerr solution was constructed, which was initially expressed in Boyer--Lindquist coordinates. In the nonrotating limit, we immediately obtain suitable coordinates for the Schwarzschild region $\Sp$. We refer to Appendix A in \cite{HennigMacedo2021} for a comparison of the two coordinate systems. Both coordinate choices are very useful for investigations near spacelike infinity and $\Scri^+$, 
and the construction of the cylinder in both approaches is geometrically equivalent.
However, the coordinates from \cite{HennigMacedo2021} are even somewhat simpler than Friedrich's coordinates. In both cases, the metric coefficients are rational functions of the coordinates, but for the representation based on Schwarzschild coordinates, these functions have lower degrees than those for the transformation starting from isotropic coordinates.
For our present discussion, we therefore choose the simpler coordinates in the region $S_+$ in Fig.~\ref{fig:ConfDiag}, while a minor modification of these provides us with suitable coordinates for the region $\Sm$.

The starting point for the conformal compactification is the Schwarzschild metric $\tilde g$ with mass $M$ in Schwarzschild coordinates $(\tilde r,\theta,\varphi,\tilde t)$,
\begin{equation}\label{eq:metricSchwCoords}
 \tilde g = \frac{\dd\tilde r^2}{1-\frac{2M}{\tilde r}}
		   +\tilde r^2\,\dd\sigma^2-\left(1-\frac{2M}{\tilde r}\right)\dd\tilde t^{\,2},\quad
 \dd\sigma^2 := \dd\theta^2+\sin^2\theta\,\dd\varphi^2.
\end{equation}
We first compactify the radial coordinate and introduce dimensionless radial and time coordinates with the transformation $(\tilde r,\tilde t)\mapsto (r,t)$,
\begin{equation}\label{eq:transformation1}
 \tilde r = \frac{2M}{r},\quad \tilde t=2M t.
\end{equation}
Next we introduce coordinates $(\rho_\pm,\tau_\pm)$ in $\Spm$ that are adapted to radial null curves,
\begin{equation}\label{eq:transformation2}
 r=\rho_\pm(1\mp\tau_\pm),\quad t=\pm\int_r^{\rho_\pm}\frac{\dd s}{F(s)},\quad
 F(s):=s^2(1-s),
\end{equation}
where we choose the upper/lower signs in $\Sp$/$\Sm$, respectively. In the new coordinates, the surface $\tilde t=t=0$ corresponds to $\tau=0$, while the cylinder is located at $\rho_\pm=0$, and $\Scri^\pm$ is at $\tau_\pm=\pm1$. Due to the zero of $F(s)$ at $s=1$, the new coordinates have a coordinate singularity at $\rho_\pm=1$, and hence we cannot include the black hole horizon. However, since we are mainly interested in the behaviour near the cylinder, this is not a problem. We consequently consider the new coordinates in a region $\rho_\pm\in[0,\rho_\mathrm{max}]$ with some positive $\rho_\mathrm{max}<1$, and $\tau_-\in[-1,0]$, $\tau_+\in[0,1]$.

The Schwarzschild metric can now be expressed as $\tilde g=\Theta^{-2}g$ with the conformal factor $\Theta=r/(2M)$ and the conformal metric
\begin{equation}\label{eq:metric}\fl
 g = \left[1\pm\tau_\pm - \rho_\pm(1\pm2\tau_\pm-\tau_\pm^2)\right]
     \frac{1\pm\tau_\pm}{\rho_\pm^2(1-\rho_\pm)^2}\,\dd\rho_\pm^2
     \mp \frac{2}{\rho_\pm(1-\rho_\pm)}\,\dd\rho_\pm\dd\tau_\pm
     +\dd\sigma^2
\end{equation}
in the two regions $\Spm$. Note that $\rho_\pm$ are null coordinates, i.e.\ curves $\rho_\pm=\mathrm{constant}$, $\theta=\mathrm{constant}$, $\varphi=\mathrm{constant}$ are null curves (and, in fact, null geodesics with respect to the conformal metric with affine parameter $\tau_\pm$).

If we consider a function $f$, like a solution to the conformally invariant wave equation, then it is useful to derive junction conditions at the surface $\tau_-=\tau_+=0$. Then we can think about solving the wave equation by first obtaining a solution $f_-$ in $\Sm$, reading off the final function and derivative values at $\tau_-=0$, and using these to obtain function and derivative values at $\tau_+=0$, which can then serve as initial data for further evolution of the function $f_+$ in the domain $S_+$. The required conditions can be obtained by studying the relation \eqref{eq:transformation2} between the coordinates $(r,t)$ and $(\rho_\pm,\tau_\pm)$ at $t=\tau_\pm=0$. 

Firstly, it follows from
$r|_{\tau_\pm=0}=\rho_\pm$ that we have $\rho_+=\rho_-$ at $\tau_\pm=0$. Hence the function values are simply related by $f_+(\rho_+,0)=f_-(\rho_-,0)$ with $\rho_+=\rho_-$. Secondly, we look at the time derivatives. Using
\begin{equation}
 \diff{f_\pm}{\tau_\pm}\Big|_{\tau_\pm=0}
 = \diff{f}{r}\diff{r}{\tau_{\pm}}+\diff{f}{t}\diff{t}{\tau_\pm}
 = \mp\rho_\pm\diff{f}{r}+\frac{\rho_\pm}{F(r)}\diff{f}{t},
\end{equation}
we initially obtain
\begin{equation}
 \diff{f_+}{\tau_+}-\diff{f_-}{\tau_-}=-2\rho_\pm\diff{f}{r} 
 \quad\mbox{at}\quad \tau_\pm=0.
\end{equation}
Next we express the $r$-derivative in terms of $\rho$-derivatives,
\begin{equation}
 \diff{f_\pm}{\rho_\pm}\Big|_{\tau_\pm=0} = \diff{f}{r}\diff{r}{\rho_\pm}
           +\diff{f}{t}\diff{t}{\rho_\pm}=\diff{f}{r}.
\end{equation}
Therefore, we obtain the following matching conditions,
\begin{equation}\label{eq:matching}
\fl
 \tau_\pm=0:\quad
 \rho_+=\rho_-,\quad
 f_+(\rho_+,0)=f_-(\rho_-,0),\quad
 \diff{f_+}{\tau_+}=\diff{f_-}{\tau_-}-2\rho_+\diff{f_-}{\rho_-}.
\end{equation}
For ease of notation, we will from now on collectively denote a function with  portions $f_\pm$ in $\Spm$ by $f$, since it will always be clear from the context to which region we are referring. Then we can think about the conditions \eqref{eq:matching} as stating that $f$ is continuous at $\tau_\pm=0$ and has a well-defined jump in the time derivative (thereby keeping in mind that this jump is required to ensure that the function with respect to the previous coordinates $(r,t)$ is continuous and differentiable at $t=0$).

Note that the ``jump term'' always vanishes at the cylinder $\rho_\pm=0$ (assuming that $f$ has a bounded $\rho$-derivative), and we simply obtain that $f$ is continuous and differentiable there. Hence the final function and derivative values in $\Sm$ can directly be taken as initial conditions in $\Sp$ for the evolution \emph{within} the cylinder, which we discuss in Sec.~\ref{sec:spacelike} below.

\section{Conformally invariant wave equation}
\subsection{Derivation}

We formulate the conformally invariant wave equation
\begin{equation}
 0= g^{ab}\nabla_a\nabla_b f-\frac{R}{6}f
 \equiv \frac{1}{\sqrt{-\det(g)}}\left(\sqrt{-\det(g)}g^{ij}f_{,i}\right)_{,j}-\frac{R}{6} f
\end{equation}
in the coordinates $(\rho_\pm,\tau_\pm,\theta,\varphi)$.  
Firstly,  we obtain the following determinant and Ricci scalar for the metric \eqref{eq:metric},
\begin{equation}
 \det(g)=-\frac{\sin^2\theta}{\rho_\pm^2(1-\rho_\pm)^2},\quad
 R=6r\equiv 6\rho_\pm(1\mp \tau_\pm).
\end{equation}
This leads to the wave equation
\begin{equation}\fl
0 = \pm2\rho_\pm(1-\rho_\pm)f_{,\rho\tau}
    +\left[\left(1\pm\tau_\pm-\rho_\pm(1\pm2\tau_\pm-\tau_\pm^2)\right)(1\mp\tau_\pm)f_{,\tau}\right]_{,\tau}
    -\bigtriangleup_\sigma f+rf
\end{equation}
in $\Spm$,
where $\bigtriangleup_\sigma f=\frac{1}{\sin\theta}(\sin\theta f_{,\theta})_{,\theta}+\frac{1}{\sin^2\theta}f_{,\varphi\varphi}$ is the angular part of the flat Laplacian.
Next we decompose the function $f$ into spherical harmonics $Y_{\ell m}$,
\begin{equation}
 f=\sum_{\ell=0}^\infty\sum_{m=-\ell}^\ell 
 \psi_{\ell m}(\rho_\pm,\tau_\pm)Y_{\ell m}(\theta,\varphi). 
\end{equation}
Using that $\bigtriangleup_\sigma Y_{\ell m}=-\ell(\ell+1)Y_{\ell m}$, we obtain an equation for the mode $\psi_{\ell m}$,
\begin{eqnarray}\label{eq:modeEQ}
\fl
0 &=& \pm2\rho_\pm(1-\rho_\pm)\psi_{\ell m,\rho\tau}
    +\left[\left(1\pm\tau_\pm-\rho_\pm(1\pm2\tau_\pm-\tau_\pm^2)\right)(1\mp\tau_\pm)
    \psi_{\ell m,\tau}\right]_{,\tau}\nonumber \\ 
 \fl   
  &&\quad   +[\ell(\ell+1)+r]\psi_{\ell m},
\end{eqnarray}
which does not explicitly depend on $m$. In the following, we will suppress the indices $\ell$ and $m$ and just refer to a mode $\psi$ for fixed $\ell$ and arbitrary $m$.

\subsection{Simple test solutions}
\label{sec:TestSols}

Similarly to the previous investigations of the conformally invariant wave equation in \cite{FrauendienerHennig2017,FrauendienerHennig2018,HennigMacedo2021}, it is useful to construct simple exact solutions, which give a first impression of the behaviour of solutions. Such solutions can most easily be derived by first constructing functions $\tilde\psi=\tilde\psi(r)$ that solve the wave equation with respect to the physical metric $\tilde g$ and are independent of the time coordinate $t$. Afterwards, we obtain the solutions $\psi=\Theta^{-1}\tilde\psi$ with respect to the conformal metric, which have a nontrivial time dependence in terms of the coordinates $(\rho_\pm,\tau_\pm)$.

The wave equation becomes particularly simple for the physical metric $\tilde g$, since the corresponding Ricci scalar vanishes, $\tilde R=0$. For time-independent functions $\tilde\psi(r)$ and some angular mode $\ell$, we obtain
\begin{equation}
 0=r^2\left((1-r)\tilde\psi_{,r}\right)_{,r}-\ell(\ell+1)\tilde\psi.
\end{equation}
The corresponding general solution $\psi$ is given by
\begin{equation}
 \psi(\rho_\pm,\tau_\pm)=\frac{c_1}{r} P_\ell\left(\frac{r-2}{r}\right)
                         +\frac{c_2}{r} Q_\ell\left(\frac{r-2}{r}\right),
\end{equation}
where $r=\rho_\pm(1\mp\tau_\pm)$ as before. Moreover, $P_\ell$ and $Q_\ell$ refer to Legendre polynomials and Legendre functions of the second kind, respectively, and $c_1$, $c_2$ are integration constants. Since $P_\ell(\frac{r-2}{r})\sim r^{-\ell}$ and 
$Q_\ell(\frac{r-2}{r})\sim r^{\ell+1}$ as $r\to0$, we choose  $c_1=0$ for regularity at the cylinder and at $\Scri^\pm$.

For the first five values of $\ell$, we obtain (for some choice of $c_2$, which is an unimportant scaling parameter)
\begin{eqnarray}
 \ell=0: &\qquad \psi = \frac1 r\ln(1-r),\\
 \ell=1: &\qquad \psi = \frac{1}{r^2}\left[(r-2)\ln(1-r)-2r\right],\\
 \ell=2: &\qquad \psi = \frac{1}{r^3}\left[(r^2-6r+6)\ln(1-r)-3r^2+6r\right],\\
 \ell=3: &\qquad \psi = \frac{1}{r^4}\Big[(r^3 - 12r^2 + 30r - 20)\ln(1 - r)\nonumber\\
		& \qquad\qquad\quad - \frac{11}{3}r^3 + 20r^2 - 20r\Big],\\
 \ell=4: &\qquad \psi = \frac{1}{r^5}\Big[(r^4 - 20r^3 + 90r^2 - 140r + 70)\ln(1 - r)\nonumber\\
		&\qquad\qquad\quad - \frac{25}{6}r^4 + \frac{130}{3}r^3 
		- 105r^2 + 70r\Big].
 \end{eqnarray}
Note that the terms in brackets have zeros at $r=0$, which compensate the singular coefficients $1/r^{\ell+1}$, and the solutions behave like $r^\ell$ as $r\to0$.

\section{Behaviour at infinity}\label{sec:infinity}

In the following, we investigate the behaviour of solutions near the cylinder and near $\Scri^-$, in order to analyse what degree of regularity the solutions can have and how this depends on our choice of initial data.

\subsection{Near spacelike infinity}\label{sec:spacelike}

We consider a mode $\psi$ near the cylinder at $\rho=0$ and expand it in the form
\begin{equation}\label{eq:expansionCyl}
 \psi(\rho,\tau)=\psi_0(\tau)+\rho\psi_1(\tau)+\rho^2\psi_2(\tau)+\dots
\end{equation}
For ease of notation, we suppress the subscript $\pm$ at the coordinates $\rho$, $\tau$, but in the following equations, it is automatically understood that  the upper sign in a '$\pm$' or '$\mp$' refers to the region $\Sp$ with coordinates $(\rho_+,\tau_+)$, and the lower sign to $\Sm$ with coordinates $(\rho_-,\tau_-)$.

Plugging this expansion into the wave equation \eqref{eq:modeEQ}, we obtain an equation for $\psi_n$,
\begin{equation}\label{eq:cylinderEQ}
 (1-\tau^2)\ddot\psi_{n}+2(\pm n-\tau)\dot\psi_{n}+\ell(\ell+1)\psi_n=R_n,
\end{equation}
where the source term $R_n$ is given by
\begin{equation}\label{eq:Rn}
 \fl
 R_n=(1\mp\tau)(1\pm2\tau-\tau^2)\ddot\psi_{n-1}
		\pm(2n-1\mp6\tau+3\tau^2)\dot\psi_{n-1}-(1\mp\tau)\psi_{n-1}
\end{equation}
for $n>0$, whereas $R_0=0$.
A dot refers to the derivative with respect to $\tau$.

We observe that, in the region $\Sp$, the left-hand side of \eqref{eq:cylinderEQ} is identical with that of the corresponding equation discussed in \cite{FrauendienerHennig2018}, even though this was derived from other coordinates, namely those that were constructed from isotropic coordinates. The source term $R_n$, on the other hand, is considerably simpler here: it only depends on \emph{one} previous function $\psi_{n-1}$ and its time-derivatives, whereas the source term in \cite{FrauendienerHennig2018} contains up to \emph{six} functions $\psi_{n-1},\dots,\psi_{n-6}$.

The equation \eqref{eq:cylinderEQ} for $\psi_n$ is an intrinsic equation at the cylinder, and it can be solved by providing appropriate initial data at $I^-$, where the cylinder and $\Scri^-$ approach each other, i.e.\ at $\tau_-=-1$. We start by investigating which of the function/derivative values of $\psi_n$ are fixed through the equation, and which can be prescribed as free initial data. This is not immediately clear due to the degeneracy of the equation at $\tau_-=-1$.

Firstly, in the limit $\tau\to-1$, we obtain from \eqref{eq:cylinderEQ} in $\Sm$ 
\begin{equation}
 \tau=-1:\quad 2(1-n)\dot\psi_n+\ell(\ell+1)\psi_n=R_n(-1)
                 \equiv 2(2-n)\dot\psi_{n-1}.
\end{equation}
Hence the derivative $\dot\psi_n(-1)$ is determined in terms of the function value $\psi_n(-1)$, unless $n=1$, where the first term vanishes.

Secondly, we consider the $k$th $\tau$-derivative $\psi_n^{(k)}$ for $k=2,3,4,\dots$
To this end, we differentiate \eqref{eq:cylinderEQ} $k-1$ times. The result has the structure
\begin{equation}\label{eq:nthderiv}
 \tau=-1:\quad 2(k-n)\psi_n^{(k)}+\mbox{lower-order derivatives}=R_n^{(k-1)}(-1).
\end{equation}
Consequently, $\psi_n^{(k)}(-1)$ is fixed in terms of lower-order derivatives at $\tau=-1$, with exception of the $n$th derivative, since the first term in \eqref{eq:nthderiv} vanishes for $k=n$.

Overall, we see that we can prescribe the function value $\psi_n(-1)$ and the $n$th derivative $\psi_n^{(n)}(-1)$. For $n\neq 0$, we can therefore choose two initial values, while for $n=0$ only the function value can be given. Due to the degeneracy of the cylinder equation \eqref{eq:cylinderEQ} at $\tau=\pm1$, the solutions can develop logarithmic singularities. In order to achieve that a solution $\psi_n$ is initially  regular at $\tau=-1$, and, furthermore, the extension into the region $\Sp$ is regular at $\tau=1$, some of the available degrees of freedom need to be chosen in a special way. These regularity conditions will be investigated in the forthcoming subsections. But first we already summarise the outcome and list in Table~\ref{tab:freepars} the parameters that can still be chosen freely for solutions that are regular at both $I^-$ and $I^+$. For that purpose, we abbreviate the function and derivative values of $\psi_n$ at $\tau=-1$ as follows,\footnote{Like the functions $\psi_n$, the quantities $\alpha_n$, $\beta_n$,\dots depend on the value of $\ell$, but for ease of notation, we do not introduce an additional index $\ell$.}
\begin{equation}\label{eq:alphabeta}
 \fl
 \alpha_n=\psi_n(-1),\quad
 \beta_n =\dot\psi_n(-1),\quad
 \gamma_n=\ddot\psi_n(-1),\quad
 \delta_n=\psi_n^{(3)}(-1),\quad
 \eps_n  =\psi_n^{(4)}(-1).
\end{equation}

\begin{table}[ht]\centering
 \begin{tabular}{l|l|l|l|l|l}
  $n$ & $\ell=0$      & $\ell=1$    & $\ell=2$   & $\ell=3$     & $\ell=4$\\
  \hline\hline
  0   & $\alpha_0$    & $\alpha_0$  & $\alpha_0$ & $\alpha_0$   & $\alpha_0$\\
  \hline
  1   & $\alpha_1$, 
        $\beta_1$     & ---		    & $\beta_1$  & ---          & $\beta_1$\\
  \hline      
  2   & $\alpha_2$, 
        $\gamma_2$    & $\alpha_2$,
                     $\gamma_2$     & ---        & $\gamma_2$   & ---\\
      & ($\alpha_1$,  
         $\beta_1$)   &             &            & ($\alpha_0$) & \\
  \hline
  3   & $\alpha_3$,
        $\delta_3$,   & $\alpha_3$,
                       $\delta_3$   & $\alpha_3$,
								    $\delta_3$   & ---          & $\delta_3$\\
      &  ($\alpha_2$) & ($\alpha_2$,
                        $\gamma_2$) &            &              & ($\beta_1)$\\
  \hline
  4   & $\alpha_4$,
        $\eps_4$      & $\alpha_4$,
						$\eps_4$    & $\alpha_4$,
						              $\eps_4$   & $\alpha_4$,
												   $\eps_4$     & ---\\
	  & ($\alpha_3$,
	     $\delta_3$)  & ($\alpha_0$,
					     $\delta_3$) & ($\beta_1$,
					                    $\delta_3$) &\\
  \hline
 \end{tabular}
 \caption{Initial data that can be chosen freely at $\tau=-1$ such that the resulting solutions $\psi_n$ are regular at $\tau=\pm1$. Parameters in brackets have been introduced at previous orders $n'<n$, where they can be chosen arbitrarily, but need to be fixed in a special way to achieve regularity at order $n$. Note that there is some freedom to choose which particular parameters are fixed and which are free. (The regularity conditions usually relate several parameters, some of which can be chosen arbitrarily, and the others are then fixed.) This table presents a possible selection of free parameters.
 }
 \label{tab:freepars}
\end{table}

For example, considering the mode $\ell=1$, we read off from Table~\ref{tab:freepars} that we can prescribe the function value $\alpha_0$ for solutions that are regular at order $n=0$. At order $n=1$, we have no further degrees of freedom, as the initial data here are completely fixed by the regularity requirement. For $n=2$, we have the full degrees of freedom and can choose the function and second-derivative values. Then, at order $n=3$, we can choose the function and third-derivative values, but now need to fix the data introduced at the previous order $n=2$. Finally, for $n=4$, we can choose the function and fourth-derivative values, but now need to fix the function value from order $n=0$ and the third-derivative value from order $n=3$.

More details about the exact regularity conditions and about which parameters are fixed by the requirement of regularity at either $I^-$ or $I^+$ are provided below. For any parameters $n$, $\ell$, the regularity conditions are obtained by constructing exact solutions to the ODE \eqref{eq:cylinderEQ} in $\Sm$ and $\Sp$ (such that the two solutions are continuously and differentiably connected at $\tau=0$, as discussed above), then expressing the integration constants in terms of the initial data and finally demanding that the coefficients of any singular terms vanish. At each order $n$, the previous solutions and their parameters determine the next source term $R_n$, and we will always assume that all regularity conditions at previous orders $n'<n$ (for the same mode $\ell$) are satisfied, i.e.\ the previous initial data are appropriately restricted when we discuss regularity at order $n$. (Otherwise, singularities from lower orders would introduce additional singularities at higher orders via singular source terms $R_n$.)

\subsubsection{Order $n=0$.}

At the lowest order $n=0$, the source term $R_0$ vanishes, and the solutions to \eqref{eq:cylinderEQ} are --- in both regions $\Spm$ --- given by
\begin{equation}
 \psi_0(\tau)=(-1)^\ell\alpha_0 P_\ell(\tau).
\end{equation}
Since these solutions are regular, no matter how we choose the initial function value $\alpha_0\equiv\psi_0(-1)$, we do not need to impose any regularity conditions. Hence we can arbitrarily choose $\alpha_0$ for each $\ell=0,1,2,\dots$

\subsubsection{Order $n=1$.}

Before we list the particular parameter values that are required for regular solutions for the first few values of $\ell$, we describe the example $\ell=1$ in more detail, in order to illustrate how we obtain the regularity conditions. 

We start by solving \eqref{eq:cylinderEQ} for $\ell=1$ in $\Sm$ to obtain
\begin{eqnarray}\label{eq:sol11}
 \Sm:\quad \psi_1(\tau) &= \alpha_1(1+\tau)\ln(1-\tau)-(\alpha_0+\alpha_1)(1+\tau)\ln(1+\tau)
 \nonumber\\
  & \quad+\frac{12\alpha_0\tau^3-C\tau^2-12(\alpha_0+2\alpha_1)\tau+C}{12(1-\tau)},
\end{eqnarray}
where $C$ is an integration constant, and a second integration constant was replaced in terms of $\alpha_1$ using the condition $\psi_1(-1)=\alpha_1$. The solution contains a term proportional to $\ln(1-\tau)$, which would be singular at $\tau=1$, but this is not relevant since the solution is only defined in $\Sm$, where $-1\le\tau\le0$. However, there is also a term $(1+\tau)\ln(1+\tau)$, which has a well-defined limit as $\tau\to-1$, but is not differentiable there. Hence, for regular solutions, we need to eliminate this term. To this end, we require that the coefficient $\alpha_0+\alpha_1$ vanishes, which is the first regularity condition. We can assume that $\alpha_0$ was already chosen and read this as a condition for $\alpha_1$. Hence we choose
\begin{equation}
 \alpha_1=-\alpha_0
\end{equation}
for solutions that are regular at $I^-$. With this choice for $\alpha_1$, we can simplify 
\eqref{eq:sol11} and eliminate $C_1$ in favour of $\beta_1$ from the condition $\dot\psi_1(-1)=\beta_1$, since the solution is now differentiable at $\tau=-1$.

The next step is to read off the values $\psi_1(0)$, $\dot\psi_1(0)$, and use these as initial data to solve \eqref{eq:cylinderEQ} in $\Sp$. The result is
\begin{eqnarray}
 \fl
 \Sp:\quad \psi_1(\tau) &= 
     \frac14(3\alpha_0-2\alpha_0\ln2-2\beta_1)(1-\tau)\ln(1-\tau)
     \nonumber\\
 \fl    
     &\quad +\frac14(2\alpha_0\ln2+\alpha_0+2\beta_1)(1-\tau)\ln(1+\tau)
     \nonumber\\
 \fl    
     &\quad +\frac{\alpha_0(2\tau^3+3\tau^2-\tau-3)
      -2(\alpha_0\ln2+\beta_1)(\tau^2-\tau-1)}{2(1+\tau)}.
\end{eqnarray}
For regularity at $\tau=1$, we need to eliminate the $(1-\tau)\ln(1-\tau)$ term. This leads to the regularity condition $3\alpha_0-2\alpha_0\ln2-2\beta_1=0$, which is easily solved by choosing
\begin{equation}
 \beta_1=\left(\frac32-\ln2\right)\alpha_0.
\end{equation}
We see that the two parameters $\alpha_1$, $\beta_1$ that were introduced at this order are both fixed by the requirement of regularity at $I^-$ and $I^+$.

In the same way, we can discuss the solutions for other $\ell$-modes. The results for the values $\ell=0,1,\dots,4$ are summarised in Table~\ref{tab:conditions1}.
\begin{table}[ht]\centering
 {\renewcommand{\arraystretch}{1.2}
  \begin{tabular}{l|l|l}
  $\ell$ & regularity at $I^-$  & regularity at $I^+$\\
  \hline\hline
  0      & ---                  & ---\\
  \hline
  1      & $\alpha_1=-\alpha_0$ & $\beta_1=\frac{3-2\ln2}{2}\alpha_0$\\
  \hline
  2     & $\alpha_1=-\alpha_0$  & ---\\
  \hline
  3     & $\alpha_1=-\alpha_0$  & $\beta_1=2(7-3\ln2)\alpha_0$\\
  \hline
  4     & $\alpha_1=-\alpha_0$  & ---\\
  \hline
 \end{tabular}}
 \caption{Parameter conditions at order $n=1$.}
 \label{tab:conditions1}
\end{table}

We see that the $\ell=0$ solutions are automatically regular at $I^\pm$ for any values of the parameters $\alpha_1$, $\beta_1$. However, for all other $\ell$, we first need to impose one condition to achieve regularity at $I^-$. In some cases, the solutions are then regular at $I^+$ as well, while in other cases an extra condition is required to achieve that. 

\subsubsection{Order $n=2, 3, 4$.}

Next we consider the orders $n=2$, $3$, and $4$. Again we solve the ODE \eqref{eq:cylinderEQ} in both regions $\Spm$, read off coefficients of singular terms and require that these vanish. This leads to the regularity conditions given in Tables~\ref{tab:conditions2}-\ref{tab:conditions4}. 

\begin{table}[ht]\centering
 {\renewcommand{\arraystretch}{1.2}
  \begin{tabular}{l|l|l}
   $\ell$ & regularity at $I^-$  & regularity at $I^+$\\
   \hline\hline
   0 & $\beta_1 = \frac12(\alpha_0+\alpha_1)$  & $\alpha_1 = -\frac14\alpha_0$\\
   \hline
   1 & ---                                     & ---\\
   \hline
   2 & $\alpha_2 = \alpha_0-\frac13\beta_1$    & 
      $\gamma_2 = \left(-3\ln^2 2+\frac{499}{35}\ln2-\frac{2819}{168}\right)\alpha_0+\left(\frac{25}{6}-2\ln2\right)\beta_1$\\
   \hline
   3 & $\alpha_2 = \left(\ln2-\frac43\right)\alpha_0$   & $\alpha_0 = 0$\\
   \hline
   4 & $\alpha_2 = \alpha_0-\frac{1}{10}\beta_1$ &
	$\gamma_2 = \left(-45\ln^2 2+\frac{38695}{154}\ln2-\frac{3623153}{9240}\right)\alpha_0$\\
	 && \quad\qquad$+\left(\frac{233}{10}-9\ln2\right)\beta_1$\\
   \hline
  \end{tabular}}
  \caption{Parameter conditions at order $n=2$.}
  \label{tab:conditions2}
\end{table}

\begin{table}[ht]\centering
 {\renewcommand{\arraystretch}{1.2}
  \begin{tabular}{l|l|l}
   $\ell$ & regularity at $I^-$  & regularity at $I^+$\\
   \hline\hline
   0 & $\alpha_2 = \frac54\alpha_0-2\gamma_2$ & ---\\
   \hline
   1 & $\alpha_2 = \left(\frac43\ln2-1\right)\alpha_0+\frac23\gamma_2$
											  & $\gamma_2 = \frac{1}{40}(11-20\ln2)\alpha_0$\\
   \hline
   2 & ---                                    & ---\\
   \hline
   3 & $\alpha_3 = -\frac{1}{15}\gamma_2$     & $\delta_3 = \frac{1}{20}
												(157-60\ln2)\gamma_2$\\
   \hline
   4 & $\alpha_3 = \left(\ln^2 2-\frac{7739}{1386}\ln2+\frac{1959953}{415800}\right)\alpha_0$                                  & $\beta_1 = -10\alpha_0\ln2+\frac{1067}{36}\alpha_0$\\
	 &  \quad\qquad$+\left(\frac15\ln2-\frac{13}{60}\right)\beta_1$
											  & \\
   \hline
  \end{tabular}}
  \caption{Parameter conditions at order $n=3$.}
  \label{tab:conditions3}
\end{table}

\begin{table}[ht]\centering
 {\renewcommand{\arraystretch}{1.2}
  \begin{tabular}{l|l|l}
   $\ell$ & regularity at $I^-$  & regularity at $I^+$\\
   \hline\hline
   0 & $\delta_3 = \frac{33}{8}\alpha_0+\frac34\alpha_3-\frac92\gamma_2$
     & $\alpha_3 = \left(\frac14\ln^2 2-\frac{65}{24}\ln2-\frac{229}{288}\right)\alpha_0$\\ 
	 && \quad\qquad$+\left(4\ln2+\frac13\right)\gamma_2$\\
   \hline
   1 & $\delta_3 = \frac12\left(\frac{93}{100}-9\ln2\right)\alpha_0-\frac32\alpha_3$
     & $\alpha_0 = 0$\\
   \hline
   2 & $\delta_3 = \left(-\frac{27}{2}\ln^2 2+\frac{4491}{70}\ln2-\frac{5097}{112}\right)\alpha_0$
        & $\beta_1 = \frac{80640\alpha_3-(80640\ln^2 2-352896\ln2+208367)\alpha_0}{128(420\ln2-289)}$\\
     & \quad\qquad $+\left(\frac{21}{2}-9\ln2\right)\beta_1+\frac{15}{2}\alpha_3$
     & \\
   \hline
   3 & --- & ---\\
   \hline
   4 & $\alpha_4 = \frac{1}{140}\left(799\ln^2 2 - \frac{33861221}{13860}\ln2 
					\right.$  
	 & $\eps_4=-4730.0886770921\alpha_0$\\
	 & \quad\qquad $+\left. \frac{2115561661}{831600}\right)\alpha_0 - \frac{1}{105}\delta_3$
	 & 		\quad\qquad $+\left(\frac{2633}{210}-4\ln2\right) \delta_3$\\
   \hline
  \end{tabular}}
  \caption{Parameter conditions at order $n=4$. Note that the first coefficient in the very last formula for $\eps_4$ can be given exactly, but only in terms of a lengthy integral. For simplicity, we only provide a numerical approximations to 10 decimal places here.}
  \label{tab:conditions4}
\end{table}

Note that the condition for regularity at $I^-$ sometimes simplifies if the condition for regularity at $I^+$ is imposed as well. For example, for $n=2$ and $\ell=3$, the second condition $\alpha_0=0$ simplifies the first condition to $\alpha_2=0$.

These considerations show that solutions $\psi_n$ on the cylinder generically have logarithmic singularities at $I^-$ and $I^+$ at most orders $n$, but we can enforce regularity by appropriately restricting the parameters $\alpha_n$, $\beta_n$, $\dots$ 

How does this relate to the characteristic initial value problem for the conformally invariant wave equation \eqref{eq:modeEQ}? In order to solve this equation, we want to prescribe initial function values of $\psi$ at the two past null hypersurfaces shown in Fig.~\ref{fig:ConfDiag}. In particular, function values $\psi(\rho,-1)$ are given at a portion of $\Scri^-$. From these, we can easily obtain the values of $\alpha_n$. Using the expansion \eqref{eq:expansionCyl} and the definition \eqref{eq:alphabeta}, we see that
\begin{equation}\label{eq:alpha1}
 \alpha_n=\frac{1}{n!}\frac{\partial^n\psi}{\partial\rho^n}(0,-1),
\end{equation}
i.e.\ these parameters can be obtained from $\rho$-derivatives of the initial data $\psi(\rho,-1)$. On the other hand, the other parameters $\beta_n$, $\gamma_n$, $\dots$ also require information about the time derivatives of $\psi$ at $\Scri^-$, since we have
\begin{equation}\label{eq:beta1}
  \beta_n=\frac{1}{n!}\frac{\partial^n\dot\psi}{\partial\rho^n}(0,-1),\quad
  \gamma_n=\frac{1}{n!}\frac{\partial^n\ddot\psi}{\partial\rho^n}(0,-1),\quad\dots  
\end{equation}
This is an interesting difference to the previous considerations for the wave equation on Schwarzschild \cite{FrauendienerHennig2017,FrauendienerHennig2018} and Kerr \cite{HennigMacedo2021} backgrounds, where initial data were provided at the Cauchy surface $\tau=0$ instead of $\Scri^-$. In these cases, all quantities appearing in the regularity conditions can directly be obtained from the initial function values and their $\rho$-derivatives. For the present characteristic initial value problem, however, we first need to study the behaviour of $\psi$ near $\Scri^-$ to find the relation between the initial function values $\psi(\rho,-1)$ and the parameters $\beta_n$, $\gamma_n$, $\dots$, which we do in the next subsection.

\subsection{Near past null infinity}\label{sec:nearScri}

Similarly to our investigation of the behaviour near the cylinder, we use an expansion of the wave function $\psi$ (for some fixed mode $\ell$) to study a vicinity of $\Scri^-$ ($\tau=-1$). Here, the expansion has the form
\begin{equation}
 \psi(\rho,\tau)=\phi_0(\rho)+(\tau+1)\phi_1(\rho)+(\tau+1)^2\phi_2(\rho)+\dots
\end{equation}
Plugging this into the wave equation \eqref{eq:modeEQ}, we obtain
\begin{equation}\label{eq:scriEQ}
 2n(1-\rho)\rho^{n+1}\left(\rho^{-n}\phi_n\right)_{,\rho}
  = (\ell+n)(\ell-n+1)\phi_{n-1} + \rho(n-1)^2\phi_{n-2},
\end{equation}
(where the source terms $\phi_{n-1}$, $\phi_{n-2}$ are, of course, only present whenever the index is nonnegative).

For $n=0$, Eq.~\eqref{eq:scriEQ} is identically satisfied. Hence there is no restriction on the function $\phi_0$ --- as expected, since $\phi_0(\rho)=\psi(\rho,-1)$ are the freely specifiable initial function values on $\Scri^-$. On the other hand, for $n>0$, we can solve \eqref{eq:scriEQ} to obtain $\phi_n$ from the previous orders $\phi_{n-1}$ and $\phi_{n-2}$. The solution for each $n$ will contain an integration constant, which is fixed  by the choice of initial data on the second past null hypersurface, cf.~Fig.~\ref{fig:ConfDiag}. If this hypersurface is located at $\rho=\rho_0$, then the function values $\psi(\rho_0,\tau)$ are given there. The initial values $\phi_n(\rho_0)$ required for solving \eqref{eq:scriEQ} can then simply be obtained from
\begin{equation}\label{eq:phiID}
 \phi_n(\rho_0)=\frac{1}{n!}\frac{\partial^n\psi}{\partial\tau^n}(\rho_0,-1),
\end{equation}
where the derivatives are $\tau$-derivatives intrinsic to the second hypersurface.
Consequently, the solution to \eqref{eq:scriEQ} explicitly reads
\begin{equation}\label{eq:iteration}
 \fl
 \phi_n(\rho) = \left(\frac{\phi_n(\rho_0)}{\rho_0^n}
   + \int_{\rho_0}^\rho\frac{(\ell+n)(\ell-n+1)\phi_{n-1}(\rho')+\rho'(n-1)^2\phi_{n-2}(\rho')}{2n(1-\rho')\rho'^{\,n+1}}\,\dd\rho'\right)\rho^n
\end{equation}
with $\phi_n(\rho_0)$ as given in \eqref{eq:phiID}.

This allows us to study the regularity of solutions on the cylinder and, in particular, at $I^\pm$. To this end, given initial data on the two past null hypersurfaces, we use \eqref{eq:iteration} to calculate the functions $\phi_1$, $\phi_2$, etc., which determine the $\tau$-derivatives of $\psi$ at $\Scri^-$. These can be used to derive the values of the parameters $\beta_n$, $\gamma_n$, $\dots$ via \eqref{eq:beta1}, while $\alpha_n$ is obtained from \eqref{eq:alpha1}. Then we can test which of the regularity conditions from Sec.~\ref{sec:spacelike} are satisfied and hence determine the degree of regularity of the solution near the cylinder. 

Moreover, since the numerical investigations in \cite{FrauendienerHennig2018,HennigMacedo2021} strongly suggest that any singularities on the cylinder also ``spread'' to $\Scri^+$ and lead to corresponding singularities there, the behaviour of solutions on the cylinder is indicative of the behaviour on $\Scri^+$ as well. It was also found that the magnitudes of these singularities exponentially decrease with the order $n$ at which they appear on the cylinder. Hence, the larger the value of $n$ for which the first singularities appear on the cylinder, the weaker are the corresponding singularities on $\Scri$.

This can be used to construct initial data for the wave equation such that the resulting solutions have any desired degree of regularity. We can simply start from a family of data that depends on a number of parameters, follow the above procedure to work out $\alpha_n$, $\beta_n$, $\dots$ and finally adjust the parameters in the initial data such that any required number of regularity conditions is satisfied.

We illustrate the analysis of the regularity of solutions for given initial data in the next section, where we study three families of initial data for the cases $\ell=0,1, 2$. 

But first we also note that an investigation of the regularity of the functions $\phi_n$ constructed via \eqref{eq:iteration} is to some extent equivalent to the regularity analysis on the cylinder, namely as far as regularity near $I^-$ is concerned. On the cylinder, near $I^-$, we found that $\phi_n(\tau)$ generally contains singular terms of the form $(1+\tau)^n\ln(1+\tau)$, which can be eliminated by choosing data subject to the above regularity conditions. Similarly, the functions $\phi_n(\rho)$ on $\Scri^-$ are found to contain terms of the form $\rho^n\ln\rho$ near $I^-$, and elimination of those is subject to satisfying conditions equivalent to the above conditions for regularity at $I^-$. 

For example, for $n=2$ and $\ell=0$, the function $\phi_2(\rho)$ can be written in terms of an integral of the initial data function $\phi_0(\rho)$. An expansion of this expression near $\rho=0$ contains a term proportional to $\rho^2\ln\rho$. The coefficient of this term vanishes if and only if the resulting parameters satisfy $\beta_1 - \frac12(\alpha_0 + \alpha_1)=0$. This is equivalent to the $\ell=0$ condition at $I^-$ in Table~\ref{tab:conditions2}.

\section{Examples}
\label{sec:Examples}

\subsection{Initial data for $\ell=0$}

We choose the initial function values $\psi(\rho,-1)=\phi_0(\rho)$ on $\Scri^-$ on an interval $0\le\rho\le\rho_0$. The second past null hypersurface is located at $\rho=\rho_0$, and we prescribe the function values $\psi(\rho_0,\tau)$ there. For the following analysis, however, it is sufficient to specify the function value and first few $\tau$-derivative values of $\psi(\rho_0,\tau)$ at $\tau=-1$. Otherwise, the function values on this second hypersurface remain arbitrary.

For $\ell=0$, we consider initial data with 
\begin{equation}\label{eq:ID0a}
 \phi_0(\rho)=4(4 - \rho)
\end{equation}
and $\psi(\rho_0,\tau)$ such that
\begin{eqnarray}\label{eq:ID0b}
 \psi(\rho_0,-1)=4(4-\rho_0),\quad
 \dot\psi(\rho_0,-1)=6\rho_0,\nonumber\\
 \ddot\psi(\rho_0,-1)=2(1+C_1)(5\rho_0-1)\rho_0,\quad
 \dddot{\psi}=3(1+C_2)(7\rho_0^2-1)\rho_0.
\end{eqnarray}
Here, $C_1$ and $C_2$ are two free constants, and we will investigate how regularity of the resulting solutions depends on the choice of these constants. (In order to explain the particular form of the previous derivative values, we note that the constants were introduced such that the most regular solutions will be obtained for $C_1=C_2=0$.) 

From the initial function values \eqref{eq:ID0a}, we calculate the first few $\alpha$-parameters using \eqref{eq:alpha1},
\begin{equation}
 \alpha_0=16,\quad
 \alpha_1=-4,\quad
 \alpha_2=\alpha_3=0.
\end{equation}
Next we obtain $\phi_1$ from \eqref{eq:iteration},
\begin{equation}
 \phi_1(\rho)=6\rho.
\end{equation}
This allows us to calculate the $\beta$-parameters from \eqref{eq:beta1},
\begin{equation}
 \beta_0=0,\quad
 \beta_1=6,\quad
 \beta_2=\beta_3=0.
\end{equation}
With these parameter values, we see that the regularity conditions for $n=2$ and $\ell=0$ are satisfied both at $I^-$ and $I^+$, cf.~Table~\ref{tab:conditions2}. Hence the function $\psi_2$ that results from these initial data will be regular everywhere on the cylinder.

In the next iteration step, we construct $\phi_2$, again using \eqref{eq:iteration},
\begin{equation}
 \phi_2(\rho) = \left(5(1+C_1) - \frac{C_1}{\rho_0}\right)\rho^2 - \rho,
\end{equation}
as well as the resulting $\gamma$-parameters,
\begin{equation}
 \gamma_0 = 0,\quad
 \gamma_1 = -2,\quad
 \gamma_2 = 10(1+C_1) - 2\frac{C_1}{\rho_0}.
\end{equation}
The parameter condition for $n=3$, $\ell=0$ (cf.~Table~\ref{tab:conditions3}) then becomes
\begin{equation}
 (5\rho_0 - 1)C_1 = 0.
\end{equation}
Hence the solution is regular at the cylinder near $I^-$ at order $n=3$ if $C_1=0$ (or in the special case $\rho_0=1/5)$, whereas otherwise a singularity is present. 
Equivalently, we can also calculate the next function $\phi_3(\rho)$ to observe that a logarithmic term proportional to $\rho^3\ln\rho$ is present whenever  $C_1\neq0$ (and $\rho_0\neq1/5$).

On the other hand, the solution is always regular near $I^+$ at order $n=3$, since no extra condition is required for that. 

If we now choose $C_1=0$, then the next function $\phi_3$ is regular and turns out to be
\begin{equation}
 \phi_3(\rho) = \frac12 \left(7(1+C_2) - \frac{C_2}{\rho_0^2}\right)\rho^3 - \frac{\rho}{2},
\end{equation}
from which we obtain the parameters
\begin{equation}
 \delta_0=0,\quad
 \delta_1=-3,\quad
 \delta_2=0,\quad
 \delta_3 = 21(1+C_2) - 3\frac{C_2}{\rho_0^2}.
\end{equation}
The conditions for regularity at order $n=4$ (cf.~Table~\ref{tab:conditions4}) then become
\begin{equation}
 I^-:\quad (7\rho_0^2 - 1)C_2 = 0,\qquad
 I^+:\quad \frac{169}{18} + \frac{10}{3}\ln2 - 4\ln^2 2  = 0.
\end{equation}
The first condition is satisfied if we choose $C_2=0$ (or in the special case $\rho_0=1/\sqrt{7}$), while the second condition is always violated. Hence we can achieve that $\psi_4(\rho)$ is regular at $I^-$, but it will always develop a logarithmic singularity at $I^+$.

In summary, we find that the family of initial data \eqref{eq:ID0a}, \eqref{eq:ID0b} leads to solutions for which the corresponding functions $\psi_n(\rho)$ are regular both at $I^-$ and $I^+$ at orders $n=0$, $1$, $2$. At order $n=3$, the solution is regular at $I^+$, while it is regular at $I^-$ only if we choose $C_1=0$ (or if $\rho_0=1/5$). If we do choose $C_1=0$, then the solution can also be regularised at order $n=4$ at $I^-$ by also choosing $C_2=0$ (or $\rho_0=1/\sqrt{7}$), but we cannot achieve regularity at $I^+$ at this order. Hence the parameters $C_1$, $C_2$ give us some control of the degree of regularity of the solutions up to the fourth order.

\subsection{Initial data for $\ell=1$}

Similarly to the previous considerations for $\ell=0$, we now analyse a family of initial data for $\ell=1$.

On $\Scri^-$, we choose the initial data
\begin{equation}\label{eq:ID1a}
 \phi_0(\rho) = 4\rho^3,
\end{equation}
and on the second past null hypersurface, we prescribe $\psi(\rho_0,\tau)$ such that
\begin{eqnarray}\label{eq:ID1b}
 \fl
 \psi(\rho_0,-1) = 4\rho_0^3,\quad
 \dot\psi(\rho_0,-1) =  -4\rho_0\left[\rho_0+\ln(1 - \rho_0)\right],\nonumber\\
 \fl
 \ddot\psi(\rho_0,-1) = -2\rho_0^2(1 + D_1)\left[\rho_0+\ln(1 - \rho_0)\right],\\
 \fl
 \dddot\psi(\rho_0,-1) = 6(1 + D_2)\rho_0^2\left[\rho_0+ 2\ln(1 - \rho_0)
    +\rho_0\ln^2(1 - \rho_0) + 2\rho_0\,\dilog(1 - \rho_0)\right]\nonumber
\end{eqnarray}
where $D_1$, $D_2$ are free parameters (again introduced such that the most regular solutions correspond to $D_1=D_2=0$), and the dilogarithm function is defined by
\begin{equation}
 \dilog(x) = \int_1^x\frac{\ln t}{1 - t}\,\dd t.
\end{equation}

From \eqref{eq:ID1a}, we obtain
\begin{equation}
 \alpha_0=\alpha_1=\alpha_2=0,\quad \alpha_3=4,
\end{equation}
and the condition for regularity at $I^-$ for $n=1$, $\ell=1$ (cf.~Table~\ref{tab:conditions1}) is identically satisfied. Consequently, we find that
\begin{equation}
 \phi_1(\rho) = -4\rho\left[\rho +\ln(1 - \rho)\right]
\end{equation}
is regular at $\rho=0$. The resulting $\beta$-parameters are
\begin{equation}
 \beta_0=\beta_1=\beta_2=0,\quad \beta_3=2.
\end{equation}
Hence the condition for regularity at $I^+$ for $n=1$, $\ell=1$ is also satisfied. Moreover, regularity for $n=2$, $\ell=1$ is then guaranteed as well, as this does not require any further condition, see Table~\ref{tab:conditions2}.

Next we calculate $\phi_2$,
\begin{equation}
 \phi_2(\rho) = -\left[\rho + \ln(1 - \rho) + D_1\Big(\rho_0 + \ln(1 - \rho_0)\Big) \right]\rho^2,
\end{equation}
and the first few $\gamma$-parameters,
\begin{equation}
 \gamma_0=\gamma_1=0,\quad
 \gamma_2=-2D_1\left[\rho_0+\ln(1 - \rho_0)\right].
\end{equation}
Plugging this into the conditions for regularity for $n=3$, $\ell=1$ (cf.~Table~\ref{tab:conditions3}), we obtain the following conditions for regularity at $I^\pm$,
\begin{equation}
 \fl
 I^-:\quad D_1\left[\rho_0+\ln(1 - \rho_0)\right]=0,\qquad
 I^+:\quad D_1\left[\rho_0+\ln(1 - \rho_0)\right]=0.
\end{equation}
Both conditions are identical and show that regularity at $I^\pm$ is equivalent to $D_1=0$ (since the term in brackets cannot vanish for $0<\rho_0<1$).

If we choose $D_1=0$, then the next function is given by
\begin{eqnarray}
 \fl
  \phi_3(\rho) &=& \rho^2
   \Bigg[D_2\frac{\rho}{\rho_0}\ln(1 - \rho_0)\left[2+\rho_0\ln(1 - \rho_0)\right] 
     + \ln(1 - \rho)\left[2+\rho\ln(1 - \rho)\right]\nonumber\\
  \fl    
  &&\qquad + \rho\left[1+D_2+2D_2\,\dilog(1 - \rho_0) + 2\,\dilog(1 - \rho) \right]\Bigg].
\end{eqnarray}
From $\phi_3$, we get
\begin{eqnarray}
 \fl
 \delta_0=\delta_1=\delta_2=0,\nonumber\\
 \fl
 \delta_3 = \frac{6D_2}{\rho_0}\ln(1 - \rho_0)\left[2+\rho_0\ln(1 - \rho_0)\right] 
  +6D_2-6 + 12D_2\,\dilog(1 - \rho_0).
\end{eqnarray}
We can use these parameters to obtain the conditions for regularity for $n=4$, $\ell=1$ (cf.~Table~\ref{tab:conditions4}). We find that the condition at $I^+$ is identically satisfied, while the condition for regularity at $I^-$ becomes
\begin{equation}
 D_2\Big[\ln(1 - \rho_0)\left[2+\rho_0\ln(1 - \rho_0)\right] + 2\rho_0\,\dilog(1 - \rho_0)  + \rho_0\Big] = 0.
\end{equation}
Hence regularity at $I^-$ can be achieved by choosing $D_2=0$ (or in the special case where the bracket vanishes, which happens for $\rho_0=0.538754875\dots$).

We see that the solution resulting from the above choice of initial data is always regular at orders $n=0, 1, 2$. If we choose $D_1=0$, then regularity at order $n=3$ is achieved both at $I^-$ and $I^+$. In that case, one of the regularity conditions at order $n=4$ is always satisfied, whereas the other condition will be satisfied for $D_2=0$.

\subsection{Initial data for $\ell=2$}
In our third and final example, we study a family of data for $\ell=2$. Here, we choose
\begin{equation}
 \phi_0(\rho) = 4\rho^2(1 - \rho)
\end{equation}
and
\begin{eqnarray}
  \psi(\rho_0,-1) = 4\rho_0^2(1 - \rho_0),\quad
  \dot\psi(\rho_0,-1) = 12\rho_0(\rho_0 - 1),\nonumber\\
  \ddot\psi(\rho_0,-1) = 2\rho_0^3+24\rho_0^2\ln2 - 50\rho_0^2 + 24\rho_0 + 2E_1,\\
  \dddot\psi(\rho_0,-1) = 12(1 + E_2)\rho_0^2\left(9\rho_0\ln2 - 13\rho_0 + 4\right),\nonumber
\end{eqnarray}
in terms of two free parameters $E_1$, $E_2$.

For these data, we obtain
\begin{equation}
 \alpha_0=\alpha_1=0,\quad
 \alpha_2=4,\quad
 \alpha_3=-4,
\end{equation}
such that the condition for regularity at $I^-$ for $n=1$, $\ell=2$ is identically satisfied (cf.~Table~\ref{tab:conditions1}).
To derive the $\beta$-parameters, we first calculate
\begin{equation}
 \phi_1(\rho)= 12\rho(\rho - 1),
\end{equation}
from which we get
\begin{equation}
 \beta_0=0,\quad
 \beta_1=12,\quad
 \beta_2=-12,\quad
 \beta_3=0.
\end{equation}
With these parameters, the second condition for $n=1$, $\ell=2$ is satisfied as well, i.e.\ the solutions are also regular at $I^+$ at this order.

For the next function $\phi_2$, we obtain
\begin{equation}
 \phi_2(\rho) = \rho^3  + \left(12\ln2 - 25 + \frac{E_1}{\rho_0^2}\right)\rho^2 + 12\rho,
\end{equation}
and the resulting $\gamma$-parameters are
\begin{equation}
 \gamma_0=0, \quad \gamma_1=24,\quad
 \gamma_2=\frac{2}{\rho_0^2}\left(12\rho_0^2\ln2 - 25\rho_0^2 + E_1\right).
\end{equation}
We can now consider the regularity conditions for $n=2$, $\ell=2$  (cf.~Table~\ref{tab:conditions2}). The condition at $I^-$ is identically satisfied, while the condition at $I^+$ becomes $E_1=0$.

We now choose $E_1=0$, such that the solution is regular everywhere on the cylinder at order $n=2$. Since no conditions are required for regularity at order $n=3$, $\ell=2$ (cf.~Table~\ref{tab:conditions3}), the next function $\phi_3$ is automatically regular as well.  We obtain
\begin{equation}
 \phi_3(\rho) = 2\left(\left(9\ln2 - 13\right)(1+E_2) + \frac{4E_2}{\rho_0} \right)\rho^3 + 8\rho^2.
\end{equation}
From $\phi_3$, we then calculate
\begin{equation}
 \delta_0=\delta_1=0,\quad
 \delta_2=48,\quad
 \delta_3 = 12\left(9\ln2-13\right)(1+ E_2)  + 48 \frac{E_2}{\rho_0}.
\end{equation}
Plugging this into the condition for regularity at $I^-$ for $n=4$, $\ell=2$ (cf.~Table~\ref{tab:conditions4}), we get
\begin{equation}
 E_2 \left(9\ln2 - 13+ \frac{4}{\rho_0}\right)=0.
\end{equation}
Hence regularity at $I^-$ can be achieved by choosing $E_2=0$ (or in the special case $\rho_0=4/(13-9\ln2)=0.591569364\dots$). However, the condition for regularity at $I^+$ is not satisfied, independently of $E_2$ and $\rho_0$, i.e.\ we cannot remove the singularity there.

In summary, we find that these solutions are always regular at orders $n=0,1$. For $n=2$, the solutions are still regular at $I^-$, and we can choose $E_1=0$ to guarantee regularity at $I^+$. Then, the solutions are automatically regular at order $n=3$ as well. Finally, at order $n=4$, we can choose $E_2=0$ for regularity at $I^-$, while regularity at $I^+$ cannot be achieved at this order.

\section{Discussion}\label{sec:discussion}

We have investigated the behaviour of solutions to the characteristic initial value problem for the conformally invariant wave equation on a Schwarzschild background. Initial data were prescribed on a portion of $\Scri^-$ and on a second null hypersurface. The evolution was then considered in a region that extends up to a portion of $\Scri^+$ and also includes the cylinder representation $I$ of spacelike infinity $i^0$.

Firstly, we observe that generic solutions are already ill-behaved near the initial hypersurface, namely at $I^-$, where $\Scri^-$ approaches the cylinder. On the one hand, this follows by studying the behaviour of the $n$th time-derivative of the solution at $\Scri^-$. These derivatives generally include logarithmic terms of the form $\rho^n\ln\rho$ in terms of a null coordinate $\rho$, which have limited regularity as the cylinder at $\rho=0$ is approached. On the other hand, the same behaviour is also revealed from a study of the  equations intrinsic to the cylinder, which describe the function values and their $n$th $\rho$-derivatives at $\rho=0$. These functions generically behave like $(1+\tau)^n\ln(1+\tau)$ in terms of the time coordinate $\tau$, and hence again show the degeneracy at $I^-$ ($\tau=-1$).

Secondly, in addition to the singularities at $I^-$, the evolution within the cylinder also reveals that further singularities tend to develop as $I^+$ at $\tau=1$ is approached, where the solutions behave like $(1-\tau)^n\ln(1-\tau)$ at order $n$. Furthermore, as we already pointed out, any singularities on $I^+$ are expected to give rise to singularities on $\Scri^+$ off the cylinder as well, which is strongly suggested by the earlier numerical studies of the Cauchy problem on Schwarzschild and Kerr backgrounds in \cite{FrauendienerHennig2018,HennigMacedo2021}.

As a side note, an interesting mathematical difference for the intrinsic cylinder equations is the required type of initial values. For the Cauchy problem as studied in \cite{FrauendienerHennig2017,FrauendienerHennig2018}, we needed to prescribe the function value and first derivative at $\tau=0$ at any order $n$. Here, however, where the evolution starts from $I^-$ at $\tau=-1$, we need to choose the function value and $n$th derivative value. As a consequence, an additional analysis of the behaviour of solutions near $\Scri^-$ was necessary, in order to extract these derivative values from the characteristic initial data, i.e.\ the function values on $\Scri^-$ and on the second null hypersurface at $\rho=\rho_0=\mathrm{constant}$.

Despite the generic singularities at $I^\pm$, we also found that a suitable fine-tuning of the initial data can avoid the singularities at an arbitrary number of orders. For that purpose, we need to ensure that the initial data satisfy some or all of the regularity conditions in Tables \ref{tab:conditions1}-\ref{tab:conditions4} (and any additional conditions if regularity beyond the cases covered by those tables is required). The easiest way to achieve the desired degree of regularity is to start from a family of initial data that depends on sufficiently many parameters, and then to adjust the parameters such that all required regularity conditions are satisfied. Three examples of such families were investigated in Sec.~\ref{sec:Examples}.

Overall, we find that the choice of regular characteristic initial data does neither guarantee nor exclude regularity of solutions on $I^+$ and $\Scri^+$ (nor even at $I^-$). Instead, the behaviour subtly depends on the choice of initial data. For very special solutions like the simple exact solutions constructed in Sec.~\ref{sec:TestSols}, there are no singularities at all. Generically, however, we observe singularities at infinitely many orders, but practically we can avoid the first few singularities through an appropriate choice of initial data. In this way, we can, for example, construct initial data that are suitable for highly-accurate numerical investigations of the characteristic initial value problem. This will be carried out in a future publication.



\end{document}